\documentclass{sig-alternate-2013}
\newfont{\mycrnotice}{ptmr8t at 7pt}
\newfont{\myconfname}{ptmri8t at 7pt}
\let\crnotice\mycrnotice%
\let\confname\myconfname%

\permission{Permission to make digital or hard copies of all or part of this work for personal or classroom use is granted without fee provided that copies are not made or distributed for profit or commercial advantage and that copies bear this notice and the full citation on the first page. Copyrights for components of this work owned by others than the author(s) must be honored. Abstracting with credit is permitted. To copy otherwise, or republish, to post on servers or to redistribute to lists, requires prior specific permission and/or a fee. Request permissions from Permissions@acm.org.}
\conferenceinfo{CIKM'15,}{October 19--23, 2015, Melbourne, Australia. \\ 
{\mycrnotice{Copyright is held by the owner/author(s). Publication rights licensed to ACM.}}}
\copyrightetc{ACM \the\acmcopyr}
\crdata{978-1-4503-3794-6/15/10\ ...\$15.00.\\
DOI: http://dx.doi.org/10.1145/2806416.2806615 }

\usepackage[ruled]{algorithm2e}
\usepackage{booktabs}
\usepackage[usenames,dvipsnames]{xcolor}
\usepackage{color}

\begin{document}

\title{Personalized Federated Search at LinkedIn}

%
%
%
%

%
%
%
%
%

\numberofauthors{3} 
%
\author{
%
%
\alignauthor Dhruv Arya \\
\email{darya@linkedin.com}
\alignauthor Viet Ha-Thuc \\
\email{vhathuc@linkedin.com}
\alignauthor Shakti Sinha \\
\email{ssinha@linkedin.com}
\and
\affaddr{LinkedIn} \\
\affaddr{2029 Stierlin Ct} \\
\affaddr{Mountain View, CA, USA}
}

\maketitle
\begin{abstract}
LinkedIn has grown to become a platform hosting diverse sources of information ranging from member profiles, jobs, professional groups, slideshows etc. Given the existence of multiple sources, when a  member issues a query like ``software engineer'', the member could look for software engineer profiles, jobs or professional groups. To tackle this problem, we exploit a data-driven approach that extracts searcher intents from their profile data and recent activities at a large scale. The intents such as job seeking, hiring, content consuming are used to construct features to personalize federated search experience. We tested the approach on the LinkedIn homepage and A/B tests show significant improvements in member engagement. As of writing this paper, the approach powers all of federated search on LinkedIn homepage.
\end{abstract}




\begin{figure}
\centering
\includegraphics[width=0.4\textwidth]{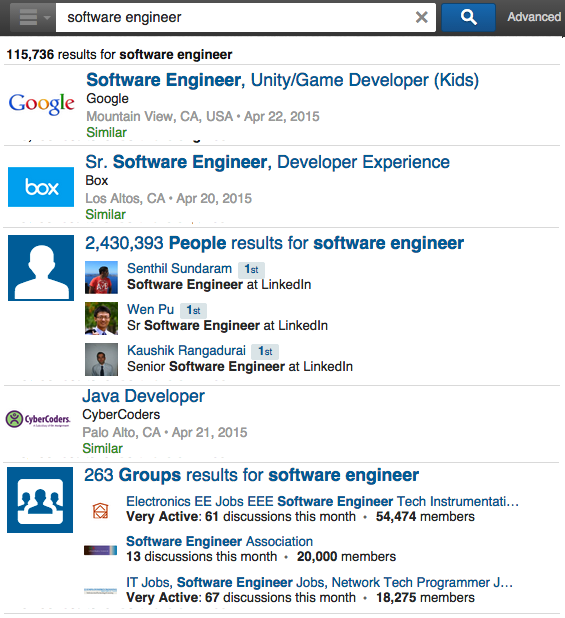}
\caption{LinkedIn Federated Search Result Page for Query ``Software Engineer''}
\label{linkedin_federated_search}
\end{figure}

\section{Introduction}
LinkedIn has grown over the years from a professional networking site to becoming a platform containing multiple professional information sources such as member profiles, jobs, professional groups, member posts and slideshows. At the same time, the member base has also increased quickly and currently has more than 350 million members. The members visit the site for various reasons ranging from searching and recruiting candidates to looking for jobs or finding professional content etc. As the number of information verticals and member base increase, the problem of serving the right information to fulfill each individual member's information need becomes more and more critical. This problem contains two subtasks including selecting the right verticals and aggregating the vertical results into a single ranking (See Figure \ref{linkedin_federated_search}) and typically is referred as federated search.


The area of federated search originated from meta search and has been actively researched in the field of information retrieval \cite{ShokouhiS11} and particularly in the context of Web search. Diaz \cite{Diaz09} addresses the problem of vertical selection, i.e., whether or not to show a specific item above the Web results given a query. Ponnuswami et. al. \cite{Ponnuswami11} and Arguello et. al. \cite{Arguello11} propose machine learning approaches for aggregating vertical results into single search result pages. They demonstrate effectiveness of the approaches on Bing search. More recently, Lefortier et. al. \cite{LefortierSRR14} present a way to blend individual vertical results and individual Web results with a case study on Yandex video search. 

However, the problem of federated search on LinkedIn presents unique challenges. First, the level of personalization in a platform like LinkedIn is much deeper than general Web search. For instance, if a member enters a query like ``software engineer'', depending on if he or she is a recruiter, job seeker or professional content consumer, the member could expect to see software engineers' profiles, jobs or slideshows on the topic. Second, individual results and blocks of results from different verticals are often associated with different features. Moreover, even if a feature is common, it is not equally important to them. This challenge is similar to federated Web search. Nonetheless, unlike Web search which typically blends \textit{either} blocks of results \cite{Arguello11}\cite{Ponnuswami11} or individual results \cite{LefortierSRR14} from different verticals, our system aggregates \textit{both} individual results (e.g. jobs as shown in Figure \ref{linkedin_federated_search}) and blocks of vertical results (e.g. people and professional group verticals). Thus, the system has to normalize features and eventually make relevance scores comparable across result verticals and result types (individual vs. block). Third, in Web search, Web results are typically the primary vertical thus they can be used as a ``pivot'' to normalize scores of the other verticals \cite{Ponnuswami11}. In our problem, the primary verticals vary depending on queries and other search context which could includes searcher's data, past activities, location etc.
 

To overcome these challenges, we propose a data-driven approach to personalize federated search. Specifically, we mine members' data and their recent activities at a large scale to understand whether or not they currently have \textit {intent} of hiring, job seeking or content consuming etc. This insight coupled with other signals are used to select verticals and aggregate vertical results into a single search result page personally relevant to each of our members. To make these signals comparable across different result categories, including verticals and result types (block vs. individual), we construct composite features combining the signals for each of the categories. Then, we let learning algorithms estimate different weights for all of the combinations (i.e., normalize the signals across the result categories) from training data.      


A/B tests done on the LinkedIn homepage shows improvements in member engagement and downstream traffic to the verticals. At the time of this writing, the work currently powers all of the federated search on the LinkedIn homepage. We organize the rest of the paper as follows. Section 2 details how we formulate federated search problem and our proposed approach. Section 3 describes searcher intent features and other signals used in the system. We discuss experimental results in Section 4. Finally, concluding remarks can be found in Section 5.

\section{Overall Approach}
\subsection{Problem Statement and Overall Framework}
Given a pair of \textit{(query, searcher)}, our task is to select from a list of all possible verticals including people (members), jobs, companies, professional groups, member posts, slideshows etc. a primary vertical and a set of secondary verticals, then rank the primary individual results and the secondary vertical blocks in a single ranked list as shown in Figure \ref{linkedin_federated_search}, without changing the order in the primary vertical. The reason for this constraint is two fold. First, we believe the base ranker of each vertical is the best one to rank results within its domain. Second, this keeps member experience consistent between federated search and vertical search. 

\begin{figure}
\centering
\includegraphics[width=0.5\textwidth]{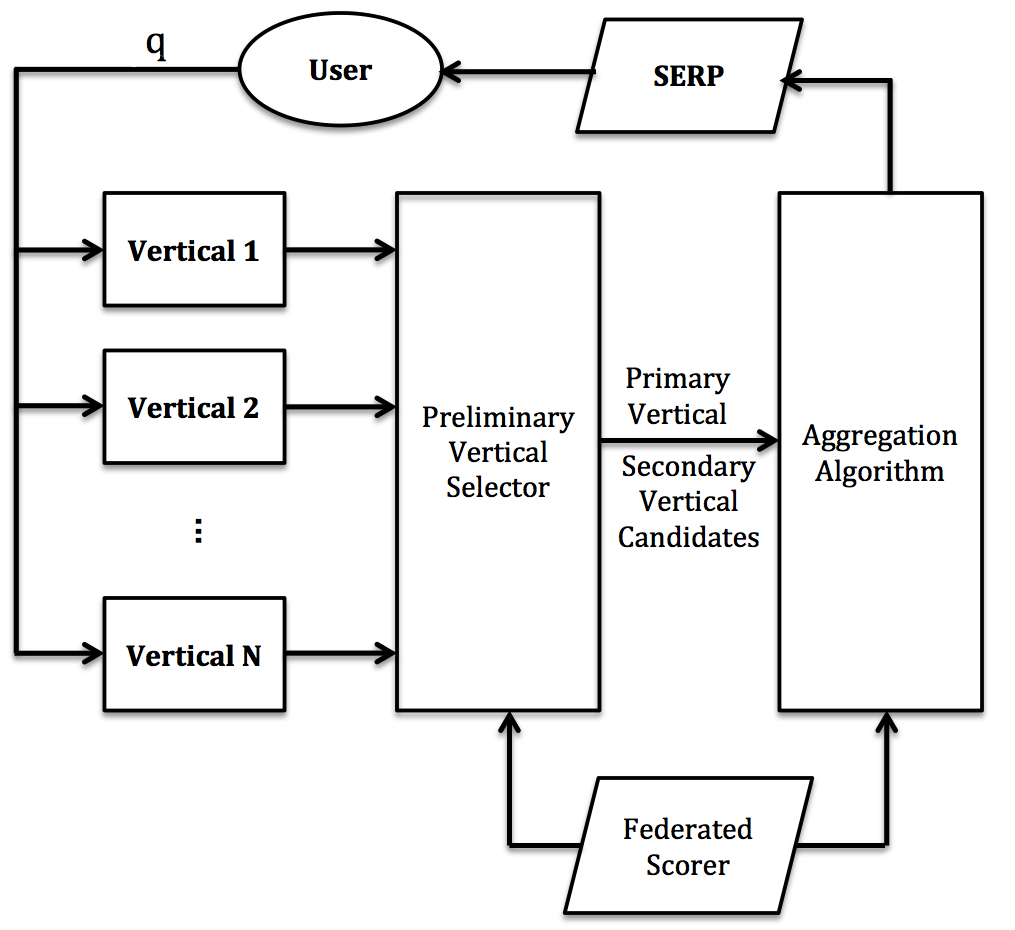}
\caption{Federated Search Overall Framework}
\label{fed_search_framework}
\end{figure}

The overall framework is described in Figure \ref{fed_search_framework}. When a member issues a query \textit{q}, the query is passed to verticals and triggers the corresponding vertical search engines to get the top \textit{K} results for each. In preliminary vertical selection phase, the federated scorer extracts features (which are described later in Section 3) and computes a relevance score for each of the verticals. The top vertical is selected as the primary one and the rest are selected as \textit{candidates} for secondary verticals. Then, in aggregation phase, these candidates compete with individual results in the primary vertical to form the final ranking. Note that these candidates are not guaranteed to show up in the ranking. Instead, depending on queries, searchers and vertical results, all, some or none of these candidates could be selected. The aggregation algorithm and the process of training the federated scorer are described in the next subsections.

\subsection{Aggregation Algorithm}
In this section we discuss how our aggregation algorithm works. Input to the algorithm includes results from the primary vertical \emph{P} along with all the candidate secondary vertical clusters \emph{C} and a federated scorer $f_s$. We go through each of primary vertical result \emph{$P_i$} computing the relevance score for this result using the federated scorer (the second loop in Algorithm 1). We compare this score with the relevance scores of all candidate secondary vertical clusters. If the former is higher, we pick the primary vertical result for $i^{th}$ position. If on the other hand there exists a candidate secondary vertical cluster that has a higher relevance score than the primary result, we add the secondary vertical cluster to the aggregated rank list and move on to the next primary result. We repeat this process till we position all the primary results. Any secondary vertical results left are dropped.

\begin{algorithm}
	\SetKwInOut{Input}{Input}
    \SetKwInOut{Output}{Output}
	
	\Input{Individual results from primary vertical \textit{P} \\
		   Secondary vertical clusters \textit{C} \\
		   Federated scorer $f_s$}
		   
    \Output{Aggregated result ranked list}

	$sortedSecondaryVerticals \rightarrow [ ]; rankList \rightarrow [ ];$

    \For{i=1 \emph{\KwTo} len(C)} {
    	\tcp{data structure sorted based on $f_s$($C_i$)}
    	sortedSecondaryVerticals.add($C_i$)
    }
    
    $j \leftarrow 1$; 
    
	\For{i=1 \emph{\KwTo} len(P)} {
		
		\eIf{$f_s$($P_i$) > $f_s$(sortedSecondaryVerticals.get(j))} {
			rankList[i] = $P_i$\\
			$i \leftarrow i + 1$\\
		}{
			rankList[i] = sortedSecondaryVerticals.get(j)\\
			sortedSecondaryVerticals.remove(j)\\
			$j \leftarrow j + 1$\\
		}
	}
	
	\KwRet{rankList}
 	
 	\caption{Federated Search Aggregation Algorithm}
\end{algorithm}

\subsection{Federated Scorer Training}
As presented above, the purpose of federated scorer is to provide a \textit{universal} relevance score for each vertical block as well as each vertical individual result. The scores have to be comparable across verticals (for preliminary vertical selection) and between vertical blocks and vertical individual results (for result aggregation). In this work, we train a federated scorer predicting probability that a vertical block or vertical individual result gets clicked if it appears on a search result page (SERP) shown to the member.

Traditionally, ground truth data is labeled by human judges. However, this approach is expensive and not scalable. Moreover, it is very hard for the judges to judge the relevance on behalf of some other member, making it challenging to apply the approach for personalized settings. Thus, in this work the training data is collected from click logs via a randomization experiment exposed to a small random fraction of LinkedIn search traffic. In this experiment, given a query we apply some business rules to come up with a few eligible verticals. We randomly pick one as a primary vertical and leave the others as secondary ones. Then, we randomly insert the secondary verticals (as blocks of results) into the primary ranking without re-ordering the primary individual results. Clicked results (either primary individual results or secondary vertical blocks) are labeled positive and skipped results (unclicked ones ranked above the last clicked result in a ranking) are labeled as negative. The results ranked below the last clicked one are dropped since it is unknown if the member ignored these results or simply did not see them. The benefit of randomization is that it avoids the bias towards the original vertical selection and ranking. Given the training data, we apply logistic regression to train a federated scorer. The features used to train the scorer are described in the next section. 

\section{Features}
\subsection{Searcher Intent}
A query can be ambiguous in the light of all information that exists in multiple verticals. For example, if a member enters a query like ``machine learning'', he or she could be interested in recruiting machine learning people, looking for jobs related to machine learning, finding professional groups on the topic to join, discovering content on the topic or some of the use cases. To tackle this issue, we take a data-driven approach to personalize search results. For instance, if we know that the member is currently looking for a job, he or she is likely to be more interested in job results than the other verticals. Similarly, if the member is hiring machine learning scientists, people results should be more important to him or her. 

\textit{Intents} of searchers such as job seeking, hiring, content consuming etc. are inferred from their profiles and past behaviors. At a high level, if a user's title is recruiter, he or she is likely to have hiring intent. Similarly, if a user is a final year student, he or she could have job seeking intent. Another source of data used to infer user intents is their recent activities. For example, if users recently searched or applied for jobs, they tend to have job seeking intent. We train a machine-learned model combining all of the signals to predict intents for all of the member base. The model is run on a daily basic to update members' intents dynamically. It is worth noting that a member could have multiple intents at the same time.       

A remaining challenge is that some evidence such as knowing a searcher has job seeking intent might be associated one or a few verticals (e.g. job vertical), but not all of them. Some evidence might be related to multiple verticals but with different levels of importance. To overcome this issue, we construct \textit{composite features}, capturing both searcher intents and result categories including both verticals and result types (individual or block). For instance, the feature below only fires if the searcher has job seeking intent and the result is a block of jobs. Otherwise, it has value of zero. We create all of the combinations of intents and result categories and learn different weights for them. In essence, we let the learning algorithm associate each of the evidence with the categories and normalize them across the categories from training data.

\[ f = \left\{ 
  \begin{array}{l l}
    1 & \quad \text{if searcher has job seeking intent} \\ 
     & \quad \quad \text{and the result is job vertical block} \\
    0 & \quad \text{otherwise}
  \end{array} \right.\]

\subsection{Keyword Intent}
The feature category aims to capture the intents (result categories) of queries. Specifically, we mine historical click logs to estimate \textit{p(result category| query)}, for instance the probability that members click on professional group vertical block for the query ``leadership''. These probabilities are computed offline for every head query and use this insight to construct the features online. One issue of this approach is that the probabilities are biased towards the previous vertical selection and ranking algorithms. Resolving this is a future direction of this work.  

\subsection{Base Ranking Features}
We also exploit information provided by vertical first pass rankers (base rankers) to construct features. One example could be relevance scores from the first pass rankers. These features also have an effect of minimizing the inconsistency between the federated scorer and the first pass rankers (recall that the order in the primary vertical is kept unchanged). One issue with this kind of information is that the relevance scores might not be calibrated well across verticals. To resolve this issue, we again construct composite features like relevance score if result is an individual job or a slideshow vertical block. For the later, we compute relevance score for each block by averaging scores of the top results in the block. Then, we let the learning algorithm normalize the scores across result categories by learning different weights for the features from the training data. 

\section{Experimental Results}
\emph{\textbf{Baseline}} is a legacy federated search algorithm at LinkedIn. It uses a set of business rules based on past member interaction with verticals associated with keywords and relevance scores returned by vertical ranking functions. It can be viewed as a  function on the feature sets described in Sections 3.2 and 3.3. Although the function is manually defined, it has been running on live traffic for a relatively long time and have been iteratively refined. The key difference between the baseline and the new approach is that the later is highly personalized by using searcher intent as a key signal.

We conducted A/B test for sufficient duration of time (6 weeks) to account for any novelty effect. A query is first tagged for existence of entities like \emph{names}, \emph{job titles}, \emph{skills} etc. As we are interested in exploratory search, we only experimented with non-name queries. Based on this condition, a random portion of LinkedIn federated search traffic from our world-wide member base is used for A/B testing. The federated search combines results from all of seven verticals available on LinkedIn including people, job, company, university, group, slideshow and members' posts. The traffic is then randomly split into control and treatment buckets. Each bucket ends up with several hundreds of thousand searches. Searches in the control bucket are processed by the baseline and the treatment bucket uses the proposed approach. We look at three metrics including primary vertical click-throught-rate (CTR), secondary vertical CTR and secondary vertical switches. The difference between the second and the third metrics is that the former is defined on clicks on \textit{individual results} within secondary clusters while the later is based on clicks on \textit{cluster headers} that take users to vertical search pages.

Table \ref{fig:metrics} shows that the proposed approach is better than the baseline on all of the metrics (all of the improvements are statistically significant). Specifically, the proposed approach is $0.62\%$ better than the baseline on primary vertical CTR. In terms of secondary vertical engagement, the proposed approach largely improves over the baseline: $4.31\%$ and $10.66\%$ improvements on secondary vertical CTR and secondary vertical switches. It is somewhat surprising that the improvement on primary vertical CTR is much lower than on secondary vertical CTR. It is probably because the proposed approach shows more relevant secondary verticals, members are more likely to switch to secondary vertical search result pages ($10.66\%$ higher). Thus, they have less chance to engage in the primary results. A deep dive into log data also reveals that the baseline tends to over-emphasize primary results and on average ranks secondary vertical clusters lower in result pages. Fully understanding and modeling the trade-off between member engagement on primary and secondary results is another future direction of this work.

%
%
%

\begin{table}
\centering
\begin{tabular}{llr}
\toprule
\textbf{Metric} & \textbf{Improvements} \\
\midrule
Primary Vertical CTR & \textcolor{ForestGreen}{\textbf{+0.62\%}} \\
Secondary Vertical CTR & \textcolor{ForestGreen}{\textbf{+4.31\%}} \\
Secondary Vertical Switches & \textcolor{ForestGreen}{\textbf{+10.66\%}} \\
\bottomrule
\end{tabular}
\caption{Metrics improvements of treatment over baseline. Due to business sensitivity, we only show relative improvements instead of absolute metric values.}
\label{fig:metrics}
\end{table}
\vfill\eject
\section{Conclusions}

In this paper, we present the problem of personalized federated search at LinkedIn and propose a data-driven approach for this problem. Our approach takes into account members' data and activities to infer their intents such as hiring and job seeking. This insight combined with other signals are used to select primary and candidate secondary verticals and then to aggregate primary individual results and the secondary clusters into a personalized ranking. Though presented in LinkedIn federated search context, the approach is applicable other domains where vertical selection and aggregation are highly personalized. Our A/B tests show that the approach could significantly improve user engagement. The approach is currently serving all of federated search on LinkedIn homepage.

One future direction of this work is to remove the bias towards the previous vertical selection and ranking algorithms in the current keyword intent features. Another direction is to understand and model the trade-off between member engagement on primary verticals and engagement on secondary verticals. Given the insight, we will determine the best balance in terms of member experience.  


%
\bibliographystyle{abbrv}
\bibliography{sigproc}  
%
%

\balancecolumns
\end{document}